\documentclass[aps,prl,reprint,twocolumn,superscriptaddress,showpacs]{revtex4}
\usepackage{graphicx}
\usepackage{amssymb,amsmath}
\usepackage{color}
\usepackage{bm}
\usepackage[final]{pdfpages}

\newcommand{\ignore}[1]{}

\begin{document}

\title{
Temporal Networks: Slowing Down Diffusion by Long Lasting Interactions}

\author{Naoki Masuda}
\affiliation{Department of Mathematical Informatics,
The University of Tokyo,
7-3-1 Hongo, Bunkyo, Tokyo 113-8656, Japan}
\author{Konstantin Klemm}
\affiliation{Bioinformatics, Institute of Computer Science,
Leipzig University, H\"{a}rtelstra{\ss}e 16-18, 04107 Leipzig, Germany}
\author{V\'ictor M. Egu\'iluz}
\affiliation{Instituto de F\'isica Interdisciplinar y Sistemas Complejos
IFISC (CSIC-UIB), E07122 Palma de Mallorca, Spain}

\date{\today}

\begin{abstract}
Interactions among units in complex systems occur in a specific sequential order
thus affecting the flow of information, the propagation of diseases, and general
dynamical processes. We investigate the Laplacian spectrum of temporal networks
and compare it with that of the corresponding aggregate network. First, we show
that the spectrum of the ensemble average of a temporal network has identical
eigenmodes but smaller eigenvalues than the aggregate networks. In large
networks without edge condensation, the expected temporal dynamics is a
time-rescaled version of the aggregate dynamics. Even for single sequential
realizations, diffusive dynamics is slower in temporal networks. These
discrepancies are due to the noncommutability of interactions.  We illustrate our
analytical findings using a simple temporal motif, larger network models and
real temporal networks.
\end{abstract}

\pacs{
89.75.Hc,
05.45.-a,
02.10.Ud 
}

\maketitle

Interactions in social, biological, and engineered networks are often being
established and dismantled in a temporal sequence rather than being static
properties. Such systems represented as networks with a sequence of time stamped
interacting node pairs are called temporal networks
\cite{HolmeSaramaki2012PhysRep}. The specifics of temporal interactions affect
accessibility \cite{Lentz2013PRL} and dynamical processes on networks such as
epidemic spreading
\cite{HolmeSaramaki2012PhysRep,Karsai2011PRE,Miritello2011PRE,Masuda2013F1000},
synchronization \cite{Fujiwara2011PRE}, random walks \cite{Starnini2012PRE}, and
consensus
\cite{FernandezGracia2011,Takaguchi2011,Baxter2011,Baronchelli2012,Maity2012}.
In the context of numerical simulations of population dynamics including
networked dynamical systems ({\it e.g.}, coupled oscillators), the comparison of
aggregate and temporal dynamics is tantamount to the choice of synchronous or
asynchronous numerical schemes for updating states of the agents. Although the
effects of the two numerical schemes have been examined, the analytical results
are scarce \cite{Nehaniv2004,Hansson2005,Cao2008IEEEAC}, and the evidence mostly
remains numerical
\cite{Huberman1993PNAS,Nowak1994PNAS,Bersini1994AIIV,Cornforth2005PD}.

Here we are interested in generic effects to which such temporal interactions may give rise. A comparison is made with the corresponding aggregate dynamics
where all interactions are present permanently. We consider dynamical systems with diffusive couplings and theoretically analyze their spectral properties, which represent various dynamics such as synchronization, random walks, and diffusive processes \cite{Donetti2006JSM,Almendral2007NJP}.
We show that diffusive dynamics is slower for the temporal network ({\it i.e.}, asynchronous update) than for the aggregate network ({\it i.e.}, synchronous update) and find qualitatively different effects, even after averaging over random temporal sequences of purely linear interactions.

{\em General framework.---}We
consider linear dynamics under a set ${\mathcal M}$ of interaction matrices.
For a time interval of length $\tau$, a matrix $M^{(0)}$ drawn from ${\mathcal M}$ determines the dynamics. Then, another matrix $M^{(1)}$ is drawn
from ${\mathcal M}$ and is active for time $\tau$, and so forth.
The $N$-dimensional state vector $\bm x(t)$ evolves according to
\begin{equation}\label{eq:linear_general}
\dot{\bm x}(t) = M^{(\lfloor t / \tau \rfloor)} \bm x(t),
\end{equation}
where $t \ge 0$. Rounding to the
next lowest integer is denoted by $\lfloor \cdot \rfloor$, so $\lfloor t / \tau \rfloor$ is the number of time intervals of length $\tau$ before time $t$.
The parameter $\tau$ measures the ratio of time scales of the dynamics of $x$ and of the evolution of the interactions.

Each specific real-world scenario of duration $r\tau$ produces a particular sequence
$S = (M^{(0)}, M^{(1)}, \ldots, M^{(r-1)})$ of interaction matrices, where
$r \in \mathbb{N}$.
The dynamics given by Eq.~\eqref{eq:linear_general} has the formal solution
$\bm x(r\tau) = T(S; \tau) \bm x(0)$ with the matrix
\begin{equation} \label{eq:t_operator_seq}
T(S; \tau) = \exp(\tau M^{(r-1)}) \cdots \exp(\tau M^{(0)})
\end{equation}
being the time evolution operator for a given sequence $S$.
An initial condition $\bm x(0)$ maps to the same $\bm x(r\tau)$ at time $r\tau$ under the dynamics with constant matrix $\left(r\tau\right)^{-1} \ln T(S;\tau)$, which we call the {\em effective} matrix of sequence $S$.

{\em Random sequences with replacement.---}Being interested in the generic effects of temporal networks, we
first consider ensembles of sequences with uniform probability, that is,
sequences generated by drawing uniformly and independently {\em with replacement} from ${\mathcal M}$.
Starting from an initial condition $\bm x(0)$, the expected state at time $\tau$ is given by
\begin{equation}
\langle \bm x(\tau) \rangle = \hat{T}(\tau) \bm x(0)
\end{equation}
with the time evolution operator now averaged over all interactions
\begin{equation}
\hat{T}(\tau) = |{\mathcal M}|^{-1} \sum_{M \in {\mathcal M}} \exp(\tau M).
\label{eq:time_op}
\end{equation}
The effective matrix for the temporal dynamics is
\begin{equation}
\hat M \equiv \tau^{-1} \ln \hat{T}(\tau).
\end{equation}

This is to be compared to the dynamics $\dot{\bm x}=M^\ast\bm x$ under
the aggregate matrix
\begin{equation}
M^\ast \equiv |{\mathcal M}|^{-1} \sum_{M \in {\mathcal M}} M,
\end{equation}
where all interactions are permanently present. The time evolution operator is given by
\begin{equation}
T^\ast(\tau) = \exp(\tau M^\ast).
\label{eq:T*}
\end{equation}
Whenever $M^\ast \neq \hat{M}$, aggregate dynamics and
temporal dynamics are different already at the level of the expectation over
random sequences of interactions. Formally, such a difference is rooted in the
fact that averaging over
interactions and integration of the dynamics do not commute in general.

{\em Edge sequences with replacement.---}Let us consider an undirected multigraph $G = (V, E)$ with nodes $V=\{1, \ldots, N\}$ and a multiset $E$ of edges given as unordered pairs of nodes. We are considering a multiset so that the same unordered pair can appear as several edges.
A single diffusive coupling between $i$ and $j$, induced by a network
edge $e=\{i,j\}$, is represented by a matrix $M^{(e)}$ with
$M^{(e)}_{ii}=M^{(e)}_{jj}=-1$, $M^{(e)}_{ij}=M^{(e)}_{ji}=+1$,
and zero at all other entries. Using a homogeneous coupling strength $\sigma$, {\it i.e.}, replacing each $M^{(e)}$ by $\sigma M^{(e)}$, amounts to a general time rescaling
$t \rightarrow t/\sigma$. Also replacing $\tau$ by $\tau / \sigma$, all resulting eigenvalues are simply scaled up by $\sigma$.
The Laplacian of the aggregate network is given by $L = -\sum_{e \in E} M^{(e)}$. Taking into account that squaring the matrix yields
$M^{(e)} M^{(e)} = -2 M^{(e)}$ \cite{directed}, the temporal evolution operator
under coupling $e$ and time $\tau$ is given by
\begin{equation}
\exp(\tau M^{(e)}) =  I + \alpha(\tau) M^{(e)},
\label{eq:L graphlet}
\end{equation}
where
\begin{equation}
\alpha(\tau)\equiv \frac{1- \exp(-2\tau)}{2}.
\label{eq:def alpha}
\end{equation}
By combining Eqs.~\eqref{eq:time_op} and \eqref{eq:L graphlet},
we obtain the ensemble averaged temporal evolution operator
for a single application of the edge as follows:
\begin{equation}
\hat{T}(\tau) = |E|^{-1} \sum_{e \in E} \exp(\tau M^{(e)}) =
I + \alpha(\tau) M^\ast,
\end{equation}
with the aggregate matrix
$M^\ast = |E|^{-1} \sum_{e \in E} M^{(e)}=-|E|^{-1} L$.
The effective interaction matrix is obtained as
\begin{equation}
\hat{M} = \tau^{-1} \ln \hat{T}(\tau) =
\tau^{-1} \ln \left[ I + \alpha(\tau) M^\ast \right].
\end{equation}

Remarkably, $\hat M$
is obtained from the aggregate matrix $M^\ast$ purely by functional calculus. Therefore, the aggregate and temporal
matrices have identical eigenspaces.
Each eigenvalue $\mu^\ast$ of the aggregate matrix $M^\ast$
maps to an eigenvalue $\hat \mu$ of the temporal matrix $\hat M$ according to
\begin{equation} \label{eq:evmapping}
\hat \mu=f(\mu^\ast,\tau) \equiv \tau^{-1} \ln
\left[1+\alpha(\tau)\mu^\ast \right].
\end{equation}
Before analyzing Eq.~\eqref{eq:evmapping} further, we remark that
$-2 \le \mu^\ast \le 0$. An eigenvalue $-2$ is obtained
if and only if $|E|=1$, {\it i.e.}, for a single interaction.
For arbitrary $\tau \neq 0$, the function $f$ has fixed points $\mu= 0$ and
$\mu = -2$. For $-2<\mu^\ast<0$, $f(\mu^\ast,\tau)$ monotonically increases with $\tau$,
and
$f(\mu^\ast,\tau)\rightarrow 0$ as $\tau \rightarrow \infty$.

In the limiting case $\tau \to 0$, temporal and aggregate dynamics coincide as $f(\mu^\ast,\tau) \to \mu^\ast$.
An increase of $\tau$ has the following consequences.
First, the dynamics slows down because eigenvalues move closer to 0.
Second, dynamical behavior qualitatively changes because the slowing down of
modes is not linear; {\it i.e.},\ the ratios between eigenvalues change. The fastest
modes are least affected when $\tau$ is increased. However, the nonlinearity
is significant only for eigenvalues of $M^\ast$ that are close to $-2$.

If $|\mu^\ast|$ is sufficiently small,
we can apply the first order
approximation $\ln (1+x) \approx x$ to
Eq.~\eqref{eq:evmapping} to obtain
\begin{equation} \label{eq:fapprox}
\hat \mu \approx \rho(\tau) \mu^{\ast},
\end{equation}
where $\rho(\tau) = (2\tau)^{-1} \left[1- \exp(-2\tau)\right]$.
To the extent that the approximation is valid for all eigenvalues of $M^\ast$, the
temporal matrix itself approximates to
$\hat{M} \approx \rho(\tau) M^\ast$.

Let us comment on the relevance of the approximation for large networks.
The eigenvalues of $M^\ast$ are lower bounded
\cite{Anderson1985Lin} by
$-\max\{ d_i+d_j : \{i,j\} \in E\}/|E|$,
where $d_i$ is the degree of node $i$, {\it i.e.},\ the number of edges incident on $i$.
For eigenvalue $\mu^\ast$, this bound translates into $|\mu^\ast| \le 2 d_\text{max} /|E|$,
where
$d_\text{max}$ is the maximum degree in the network. The approximation
given by Eq.~\eqref{eq:fapprox} improves for a growing network size if $d_\text{max}/|E|
\rightarrow 0$; {\it i.e.},\ the fraction
of edges owned by each node tends to zero. Then, all eigenvalues of $M^\ast$
tend to zero, and the expected temporal dynamics is just a time-rescaled
version of the aggregate dynamics.

Otherwise, a node accumulates a finite fraction of the edges with growing $N$,
a phenomenon called Bose-Einstein condensation in
networks \cite{Bianconi2001PRL}.
Then, an eigenvalue $\mu^\ast$ of $M^\ast$ may remain finitely separated from
zero even when $N \rightarrow \infty$.

The largest deviation from the approximation is obtained when the fastest
eigenvalue of the aggregate matrix $M^\ast$ remains at $-1$ \cite{lbound} while the slowest eigenvalue
tends to zero when $N \rightarrow \infty$. In this case, the slow and fast modes are decelerated by the factors $\rho(\tau)$ and
$f(-1,\tau)$, respectively, when passing from the aggregate to temporal dynamics.
The ratio between the deceleration factors is maximum in the limit
of large $\tau$, where we obtain
\begin{equation}
\lim_{\tau \rightarrow \infty}  \frac{\rho(\tau)}{f(-1,\tau)}
=
\frac{-1}{2 \ln (1/2)} \approx{0.72}.
\end{equation}
The slow modes
are decelerated less than fast modes, and the ratio between the eigenvalues corresponding to the slow and fast modes does not fall below 72\% of the
original value. The bound is attained in the case of a star, which has the Laplacian eigenvalues
$0,1$, and $N$, translating into eigenvalues $0$, $\mu^\ast_\text{slow}=-1/(N-1)$, and $\mu^\ast_\text{fast} = - N/(N-1)$ for matrix $M^\ast$.
Networks possessing a large ratio
$\mu^\ast_\text{fast}/\mu^\ast_\text{slow}$, including scale-free networks,
are known to be difficult to synchronize in coupled chaotic dynamics \cite{Nishikawa2003PRL,Donetti2006JSM}.
Although $\mu^\ast_\text{fast}$ does not remain $O(1)$ as $N\to\infty$, such networks deviate from Eq.~\eqref{eq:fapprox} more than homogeneous networks of the same size do.

\begin{figure}
\centerline{\includegraphics[width=0.45\textwidth]{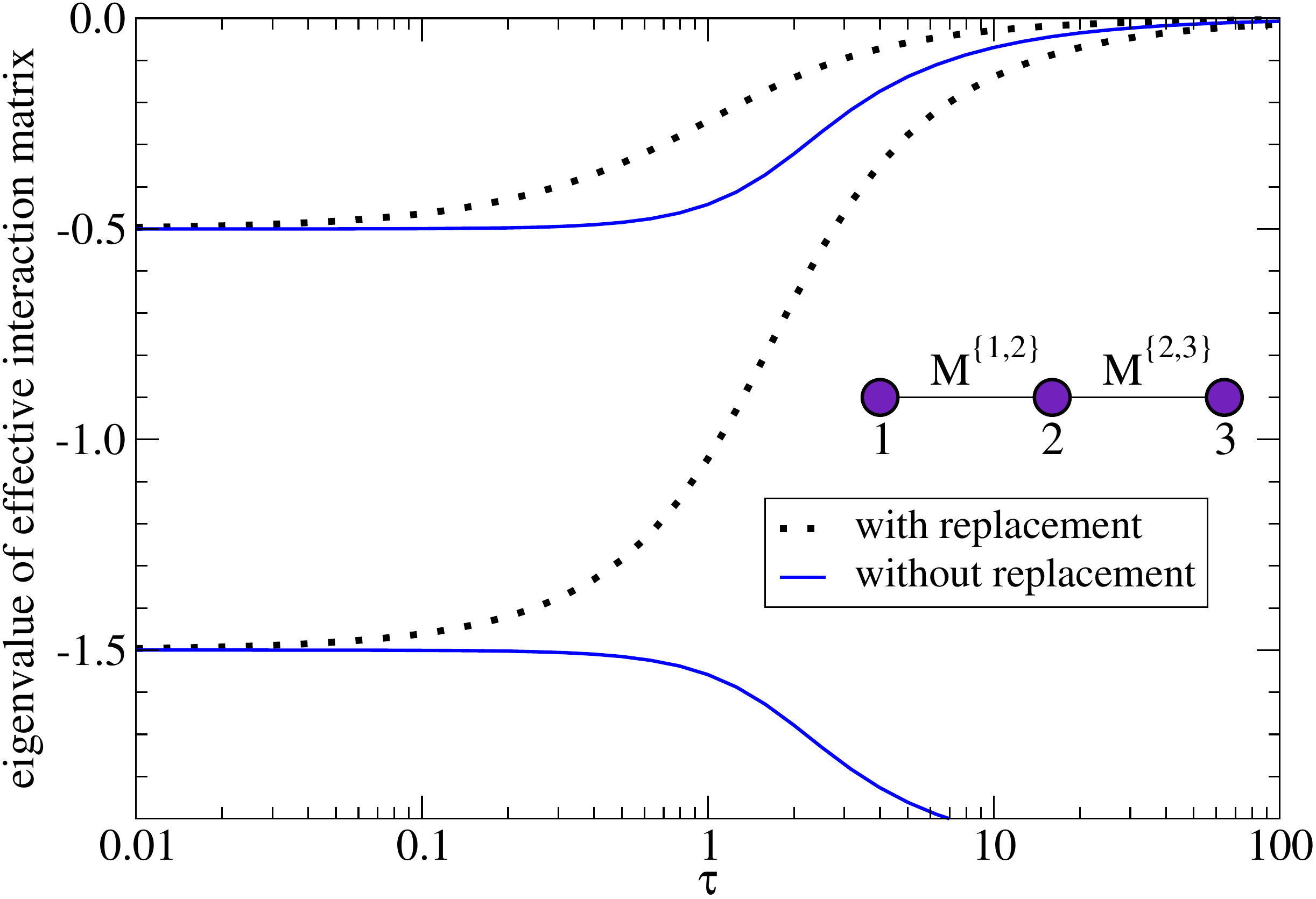}}
\caption{\label{fig:eigen_illu_temp}(color online)
   Dependence of the spectrum of effective interaction matrices on the
   temporal parameter $\tau$. We show the two nonzero eigenvalues of
   effective interaction matrices for a network with two edges
   connecting three nodes. Dotted curves: Random sequences with
   replacement. Solid curves: Random sequences without replacement.
}
\end{figure}

{\em Small example.---}For illustration, we consider a network with $N=3$ nodes and two edges,
$E=\{\{1,2\},\{2,3\}\}$. The eigenvalues of the aggregate matrix
are equal to $\mu^\ast=0$, $-1/2$,
and $-3/2$.  For the temporal dynamics averaged over random sequences with replacement, Eq.~\eqref{eq:evmapping} results in $\hat\mu = 0$, $f(-1/2,\tau)$, and $f(-3/2,\tau)$.
As the theory predicts, $\hat\mu$ (dotted curves in Fig.~\ref{fig:eigen_illu_temp}) is closer to zero than $\mu^\ast$ is, and $\hat\mu$ approaches zero as $\tau$ increases. In particular, the spectral gap,  {\it
i.e.}, the eigenvalue with the smallest nonzero absolute value, is smaller for the temporal than for the aggregate dynamics; the spectral gap gives the time scale of relaxation.

If we select
$S=(M^{\{1,2\}},M^{\{1,2\}})$ or $S=(M^{\{2,3\}},M^{\{2,3\}})$ in the sampling with replacement, the spectral
gap trivially vanishes. For sequences $S=(M^{\{1,2\}},M^{\{2,3\}})$ and
$S=(M^{\{2,3\}},M^{\{1,2\}})$, the eigenvalues of the effective matrix are
plotted by the solid curves in Fig.~\ref{fig:eigen_illu_temp}.
Also in this sampling scheme, in which we avoid multiple sampling of the same edge and the resulting loss of connectivity of the entire network, the spectral gap, at least, is smaller for the temporal than for the aggregate dynamics.

{\em Edge sequences without replacement.---}We now consider edge sequences sampled without replacement.
In the previous example, it corresponds to the two
sequences containing both edges (solid curves in Fig.~\ref{fig:eigen_illu_temp}).
For a general network and an arbitrary permutation $S = (M^{(0)}, M^{(1)}, \ldots, M^{(|E|-1)})$ of its edges,
the spectral gap of the effective temporal matrix in this case is
smaller than or equal to that of the aggregate matrix.

To show this, we
use Eqs.~\eqref{eq:L graphlet} and \eqref{eq:def alpha}
to write the time evolution operator according to Eq.~\eqref{eq:t_operator_seq} as
\begin{equation}
T(S;\tau) = [I + \alpha(\tau) M^{(|E|-1)} ] \cdots
[I + \alpha(\tau) M^{(0)} ].
\label{eq:evolution operator without replacement}
\end{equation}
If $\phi$ is a so-called spectral
function, $\phi(e^A e^B) \ge |\phi(e^{A+B})|$ holds true for
general matrices $A$ and $B$ \cite{Cohen1982LAA,Bebiano2004LAA}. The product of
the two largest eigenvalues is a spectral function.
Because the largest
eigenvalue of the evolution operator for the Laplacian dynamics
is equal to unity for both aggregate and
temporal dynamics, with the right eigenvector being
$(1\; \cdots \; 1)^{\top}$, the second largest eigenvalue of
$e^{M^{(i)}} e^{M^{(j)}}$ is at least
that of $e^{M^{(i)}+M^{(j)}}$ for arbitrary $i$ and $j$.
Therefore, in terms of the absolute value,
the spectral gap for the effective matrix
$(|E|\tau)^{-1}\ln T(S;\tau)$ is at most $|\mu^\ast|$.
This result implies that
the dynamics on an arbitrary temporal network is slower than (or at least not faster than) that on the aggregate network.

When $\tau\to 0$, we can approximate the right-hand side of Eq.~\eqref{eq:evolution operator without replacement} using $\alpha(\tau)\approx \tau$ by
\begin{equation}
I + \alpha(\tau)\sum_{i=1}^{|E|} M^{(i)}
= I+\tau|E|M^\ast \approx T^\ast(|E|\tau).
\end{equation}
Therefore, the temporal dynamics approaches the aggregate dynamics as $\tau\to 0$ as in
the case of the random sampling with replacement.

\begin{figure}
\begin{center}
\includegraphics[width=0.49\textwidth]{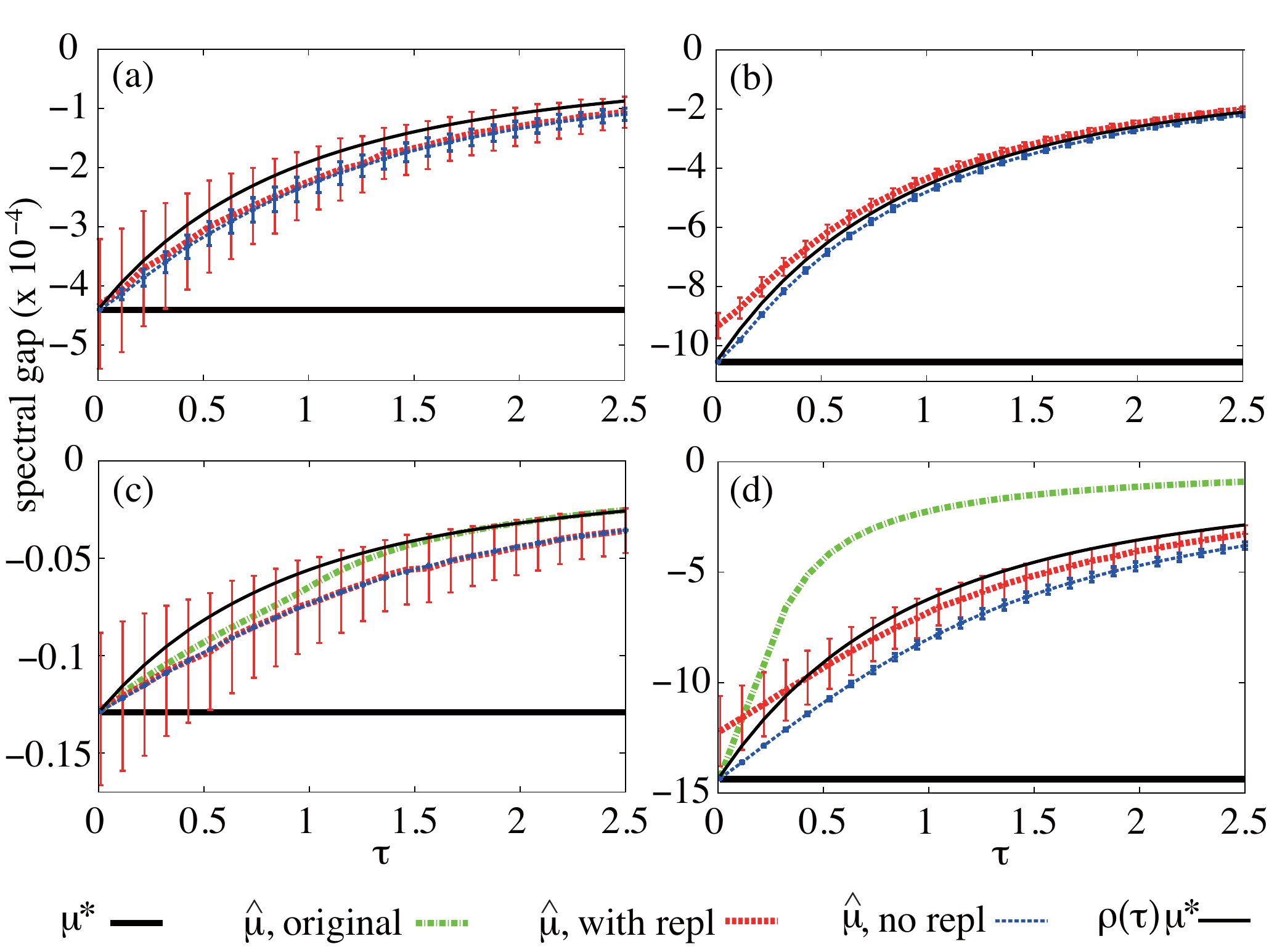}
\caption{
(color online) The (negative) spectral gap of the aggregate and temporal
  Laplacian dynamics for four networks. The error bars indicate the standard deviation calculated on the basis of $10^3$ realizations of edge sequences.
(a) Results for a random temporal graph having $N=500$ nodes and $|E|=19780$ edges.
(b) Results for a
scale-free temporal network with $N=500$ and $|E|=19990$.
We set the
parameters of the Barab\'{a}si-Albert model to $m=m_0=4$.
The degree
$d$ obeys a power law with exponent $-3$ \cite{Barabasi1999Sci}.
(c) Results for human interaction data
with $N=104$ and $|E|=782682$ among students, staff, and faculty members at the Massachusetts Institute of Technology. The data were produced by the Reality Mining Project \cite{Eagle2006PUC}.
Although the original network contains 106 subjects,
we used the largest connected component containing 104 subjects.
(d) Results for the human interaction data during a conference recorded in the SocioPatterns Project
\cite{Isella2011JTB}. Although the original data have
113 nodes and 20818 edges, we excluded one node and the link
emanating from this node, possessing two edges on it before the analysis.
This is because the spectral gap value is very
sensitive to such a nearly isolated
node. The modified network has $N=112$ and $|E|=20816$.}
\label{fig:spectral gap}
\end{center}
\end{figure}

{\em Application to model and real networks.---}For two model temporal networks and two real temporal networks,
the (negative) spectral gaps $\mu^\ast$ and $\hat \mu$ are compared in
Fig.~\ref{fig:spectral gap}. For random sampling without replacement, we appropriately normalized the spectral gap of $\tau^{-1}\ln T(S;\tau)$ by dividing by $|E|$.

The first model network is generated on a realization of
the Erd\H{o}s-R\'{e}nyi random
graph having 500 nodes and 1978 edges, which is the aggregate network. Because adjacent nodes usually
have multiple edges in real data \cite{HolmeSaramaki2012PhysRep}, we assumed 10 edges between each
pair of adjacent nodes in the aggregate random graph. Figure~\ref{fig:spectral
  gap}(a) indicates that, under the sampling both with and without
replacement, $|\hat \mu|$ i.e., the spectral gap for the temporal network averaged over $1000$ sequences (thick and thin dashed lines),
 is consistently smaller than $|\mu^\ast|$ (thick solid line)
for any $\tau$. The results are qualitatively the same for a scale-free
temporal network
[Fig.~\ref{fig:spectral gap}(b)] constructed by placing 10 edges on
each edge of an aggregated network, which is a realization of the Barab\'{a}si-Albert model
\cite{Barabasi1999Sci} having 500 nodes and 1990 edges.
The results for human interaction data obtained from the Reality Mining
Project, having 104 nodes and 782682 edges \cite{Eagle2006PUC}, and
those from the SocioPatterns Project,
having 112 nodes and 20816 edges
\cite{Isella2011JTB}, are shown in
Figs.~\ref{fig:spectral gap}(c)
and \ref{fig:spectral gap}(d), respectively.

First of all, $\mu^\ast$ is independent of $\tau$ by definition (thick solid lines).
For all the networks,
we verified that
$|\hat\mu| < |\mu^\ast|$ holds
true for the temporal networks derived from the real-world interaction sequences [dashed-dotted lines labeled ``original'' in
Figs.~\ref{fig:spectral gap}(c)
and \ref{fig:spectral gap}(d)] and
individual temporal networks generated by
the sampling without replacement (ensemble averaged
values are shown by the thin dashed lines). It should be noted that it is not the case for the
sampling with replacement (averages are shown by
the thick dashed lines).
Second, for both the sampling with replacement and without replacement,
Fig.~\ref{fig:spectral gap} indicates that
$|\hat\mu| < |\mu^\ast|$ holds true on the average and that
$|\hat\mu|$ decays toward zero as $\tau$ increases.
Third, the original sequence of edges and random sequences both with and without replacement yield values of $\hat \mu$ approaching $\mu^\ast$ as $\tau\to 0$.
All these numerical results are consistent with our theoretical results.
Finally, Eq.~\eqref{eq:fapprox} suggests that $\hat\mu$
in the case of sampling with replacement is approximated by
$\rho(\tau)\mu^\ast$. Figure~\ref{fig:spectral
  gap} shows that this is a reasonable approximation (thin solid lines).

The error bars in
Fig.~\ref{fig:spectral gap} are large in some cases, indicating that
the $\hat \mu$ value depends much on the individual sequence. This is also implied by the discrepancy between the  $\hat \mu$ values
for the real temporal networks
[dashed-dotted lines in Figs.~\ref{fig:spectral gap}(c) and \ref{fig:spectral gap}(d)] and those averaged over samples (thick and thin dashed lines).
When any pair of $M^{(i)}$ ($1\le i\le |E|$) commutes, the time evolution operators for the aggregate and temporal dynamics [Eqs.~\eqref{eq:T*} and \eqref{eq:evolution operator without replacement}]
are identical.
This is tantamount to saying that the pairwise noncommutability of graphlets may be a main source of the discrepancy between $\hat \mu$ and $\mu^\ast$ and also that between $\hat \mu$ for different temporal networks sharing the same aggregate network.

{\em Discussion.---}We have quantitatively shown that the effect of
temporal networks on diffusive dynamics is considerable. Our results imply that
synchronization is more difficult in temporal than in the corresponding
aggregate networks for general $\tau > 0$, in agreement with the numerical
results in Ref.~\cite{Fujiwara2011PRE}. The slowing down due to temporal
dynamics is also observed in other diffusive dynamics such as random walks
\cite{Starnini2012PRE}, the voter model
\cite{FernandezGracia2011,Takaguchi2011}, and the naming game
\cite{Baronchelli2012}.
Our theoretical results also enlarge
previous theoretical understanding of
synchronization dynamics in temporal networks.
The theoretical results obtained under the
framework of switching topology mostly treat the case of fast
switching, which corresponds to $\tau\to 0$ in the present study
\cite{Bekylh2004PhysicaD,Stilwell2006SIAM,So2008Chaos,Frasca2008PRL}.
In the limit $\tau\to 0$, the stability of
the aggregate Laplacian matrix is theoretically sufficient for
synchronization in temporal networks
\cite{Bekylh2004PhysicaD,Stilwell2006SIAM,So2008Chaos,Frasca2008PRL},
which is consistent with our results.
We obtained a quantitative theory for general $\tau$
without strong constraints like simultaneous diagonalizability of
the graphlets \cite{Zhao2009Auto}.
Going beyond purely diffusive dynamics, the spectral framework is relevant for
systems with local dynamics at each node. A stability criterion for the
synchrony of such dynamics across nodes is based on the eigenvalues of the
coupling matrix \cite{Pecora1998PRL}. Tests with simple systems of chaotic
oscillators provide the first evidence that the present spectral analysis for
time-dependent coupling matrices correctly combines with that approach (see
the Supplemental Material \cite{SM}).
Finally, real interaction sequences show correlated patterns \cite{Eckmann2004PNAS,Miritello2011PRE}. To reveal the relationship between
such correlated patterns and the behavior of
individual temporal sequences is warranted for future work.

We thank the SocioPatterns collaboration (http://www.sociopatterns.org) for providing the data set. We also acknowledge financial support provided by
Grants-in-Aid for Scientific Research (No.~23681033) from MEXT, Japan,
the Nakajima Foundation, VolkswagenStiftung, and MINECO (Spain) and FEDER (EU)
through the MODASS project (No. FIS2011-24785).

\bibliography{temp_nets}

\onecolumngrid

 \ \\ \ \\ \ \\ \ \\ \ \\ \ \\ \ \\ \ \\ 
 \ \\ \ \\ \ \\ \ \\ \ \\ \ \\ \ \\ \ \\
 \ \\ \ \\ \ \\ \ \\ \ \\ \ \\ \ \\ \ \\ 
 \ \\ \ \\ \ \\ \ \\ \ \\ \ \\ \ \\ \ \\
 \ \\ \ \\ \ \\ \ \\ \ \\ \ \\ \ \\ \ \\ 



\section*{Supplementary material}

# Temporal networks: slowing down diffusion by long lasting interactions

Naoki Masuda,[1] Konstantin Klemm,[2] and Víctor M. Eguíluz[3]

[1]*Department of Mathematical Informatics, The University of Tokyo, 7-3-1 Hongi, Bunko, Tokyo 113-8656, Japan*
[2]*Bioinformatics, Institute of Computer Science, Leipzig University, Härestraíße 16-18, 04107 Leipzig, Germany*
[3]*Instituto de Física Interdisciplinar y Sistemas Complejos IFISC (CSIC-UIB), E07122 Palma de Mallorca, Spain*
(Dated: November 1, 2013)

We broaden the developed framework by considering reaction-diffusion systems on temporal networks. The reaction part of the system is represented by one or more degrees of freedom assigned to the nodes and subject to dynamical rules. For simplicity, the local dynamics will be assumed to be the same at each node. Between nodes, degrees of freedom are diffusively coupled.

We consider non-linear dynamics with several coupled degrees of freedom at each node. Specifically, we study chaotic Rössler oscillators [1] with three degrees of freedom $(x, y, z)$. The evolution of the oscillator at node $i$ is given by

$$\dot{x}_i(t) = -y_i(t) - z_i(t) + \sum_{j=1}^{N} M_{ij}^{(\lfloor t/\tau \rfloor)} x_j(t) \quad \text{(s1)}$$

$$\dot{y}_i(t) = x_i(t) + 0.2 y_i(t) \quad \text{(s2)}$$

$$\dot{z}_i(t) = 0.2 + z_i(t)(x_i(t) - 7.0) . \quad \text{(s3)}$$

The sum in the equation for $x$ is a time-dependent diffusive coupling with matrix $M^{(t)}$, analogous to that in Eq. (1) in the main text. Generalizing the framework, we introduce a tunable strength $\sigma > 0$ for the diffusive coupling. Now $M_{ii}^{(e)} = M_{jj}^{(e)} = -\sigma$ and $M_{ij}^{(e)} = M_{ji}^{(e)} = +\sigma$ are the non-zero entries of the coupling matrix of edge $e = \{i, j\}$.

How does the additional parameter $\sigma$ modify the spectrum of the effective coupling matrix? Compared to the main text (Eq. (10) there), the time evolution operator for time $\tau$ is formally the same but now to be evaluated at $\tau\sigma$ so we find

$$\hat{T}(\tau\sigma) = I + \alpha(\tau\sigma)M^* . \quad \text{(s4)}$$

The effective coupling matrix is

$$\hat{M} = \tau^{-1} \ln[I + \alpha(\tau\sigma)M^*] . \quad \text{(s5)}$$

An eigenvalue $\mu^*$ of $M^*$ maps to

$$\hat{\mu} = \tau^{-1} \ln\left[1 + \frac{1 - \exp(-2\tau\sigma)}{2}\mu^*\right] . \quad \text{(s6)}$$

A connection between the eigenvalues of the coupling matrix and the stability of synchrony is obtained with the master stability approach [2]. In a linear stability criterion, the synchronous mode is stable if and only if all eigenvalues of the effective coupling matrix lie in a given

interval. Too weak coupling naturally leads to instability of the synchronous solution. On the other hand, also too strong coupling makes synchrony unstable in the Rössler system above. Specifically, all real eigenvalues of the coupling matrix need to be above $\lambda_{\mathrm{drum}} \approx -4.5$ to obtain stability, see Figure 1 in Ref. [2]. It should be noted that the sign of eigenvalues is inverted because we are working with an additive rather than subtractive coupling here.

The theory has been developed for time-independent diffusive coupling matrices. Here we ask if the prediction of stable synchrony may be based in the same way on the eigenvalues of the effective matrix in the system with non-constant *temporal* interactions between the nonlinear oscillators. For star networks of two different sizes, Figure S1(a) shows that the transition between stability and instability is shifted to larger values of the coupling strength $\sigma$ when increasing $\tau$. For a star with $N = 3$, a transition to stable synchrony is observed when the coupling strength $\sigma$ passes the value 3.0 from above in case $\tau = 0$ (aggregate dynamics). Under temporal coupling with $\tau = 0.2$, however, the onset of stability (2%) is observed already at $\sigma \approx 3.9$. Both combinations $(0, 3.0)$ and $(0.2, 3.9)$ for parameters $(\tau, \sigma)$ generate an eigenvalue $\hat{\mu} = \lambda_{\mathrm{drum}} = -4.5$ in the effective coupling matrix.

In panel (b) of Figure S1, we replot the same numerical results as a function of the eigenvalue $\hat{\mu}$ of the effective coupling matrix. We see that the stability threshold is indeed obtained by the eigenvalue $\hat{\mu}$ in good approximation. There is, however, not a collapse of the whole stability curves for different values of $\tau$. An increase of $\tau$ does not just shift the relevant eigenvalue (as an average effect) but also brings about larger fluctuations across stochastic realizations. Therefore the transition is discontinuous only for $\tau = 0.0$ and becomes smoother with larger $\tau$.

Summarizing this Supplemental Material, the eigenvalues of the effective coupling matrix provide the correct input to the master stability approach. Thus the spectral analysis of the effective matrix has bearings on the stability of non-linear dynamics of diffusively coupled elements. As shown there, temporal dynamics tends to slow down diffusion as compared to the aggregate dynamics. Thus intuitively, when going from aggregate to temporal network dynamics, larger coupling strength is required for the onset of stable synchrony. We have checked numerically that this is the case in the above scenarios (results

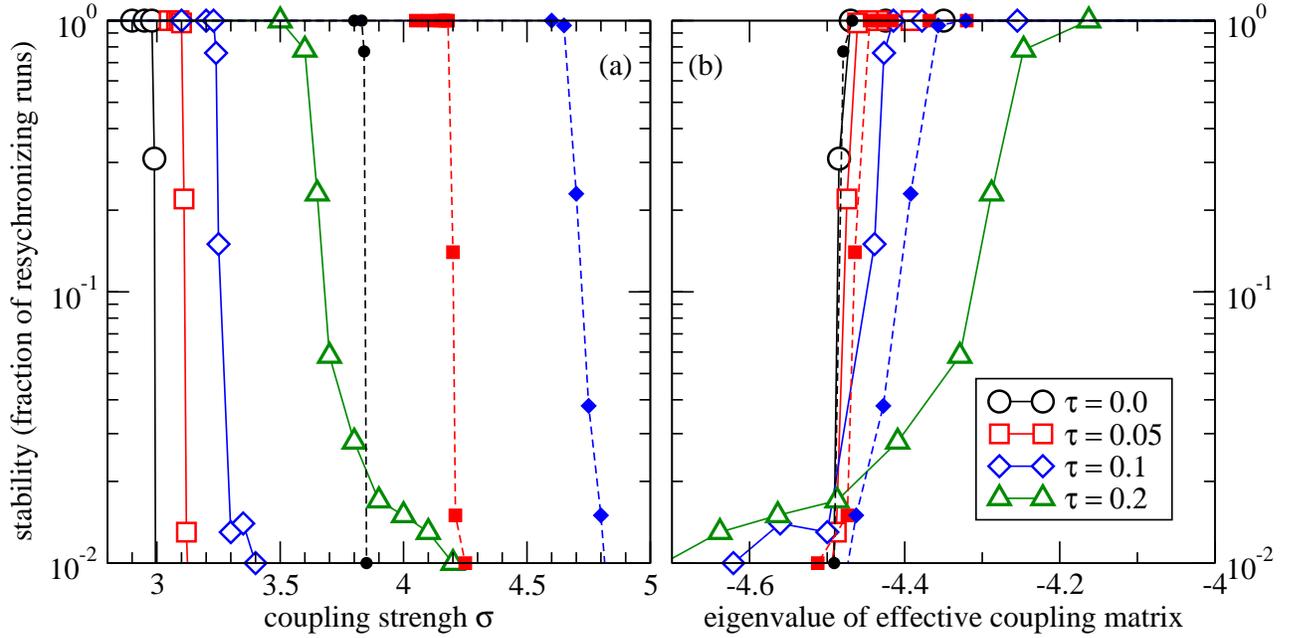

FIG. S1: (a) Stability of synchronization of local growth dynamics under varying coupling strength $\sigma$. (b) Plotting the same values as a function of the eigenvalue $\tilde{\mu}$ transformed by Eq. (s6) in this Supplementary Material. Master stability theory [2] predicts the transition to occur at $\tilde{\mu} = -4.5$. Networks are star graphs with $N = 3$ nodes (large open symbols) and $N = 7$ nodes (small filled symbols). The relevant eigenvalue (the largest in absolute value) of the aggregate coupling matrix is $\mu^* = N/(N-1)$. Each data point is the fraction of resynchronizing runs out of 1000 independent runs. A run is performed as follows. Draw the initial state of node 1, $(x_1(0), y_1(0), z_1(0))$, as three independent standard random numbers; copy the state of node 1 to all other nodes. Run the dynamics until time $t = 100$; perform a perturbation by adding to $x_i(100)$ a random number drawn uniformly from $[0, 10^{-4}]$ independently for each node $i$; run the dynamics until the variance of $x$-variables across nodes is either (i) below $10^{-13}$ or (ii) above 1.0 for a time interval of length at least 1. In case (i), the dynamical run is called resynchronizing.

not shown). The results here, however, go beyond this effect. The loss of stability at too large coupling strength is described by our theory as well. In this case, the eigenvalue of largest magnitude, rather than the spectral gap, is the quantity relevant for the stability threshold.


\end{document}